\documentclass[amssymb,amsmath,superscriptaddress,footinbib]{revtex4}
\usepackage{natbib,graphicx,subfigure,subeqnarray,fancyhdr,amsmath,multirow,color, mathrsfs}
\begin{document}
\renewcommand{\labelitemi}{$-$}

\newcommand{\Fc}{\mathcal{F}}\newcommand{\Rc}{\mathcal{R}}\newcommand{\dd}{\mathrm{d}}
\newcommand{\ee}{\mathrm{e}}\newcommand{\ci}{\mathrm{i}}\newcommand{\ib}{\mathbf{i}}
\newcommand{\jb}{\mathbf{j}}\newcommand{\kb}{\mathbf{k}}\newcommand{\ab}{\mathbf{a}}
\newcommand{\Fb}{\mathbf{F}}\newcommand{\fb}{\mathbf{f}}\newcommand{\Gb}{\mathbf{G}}
\newcommand{\Mb}{\mathbf{M}Ä}\newcommand{\nb}{\mathbf{n}}\newcommand{\Sb}{\mathbf{S}}
\newcommand{\Sbs}{\mathbf{S^*}}\newcommand{\Rb}{\mathbf{R}}\newcommand{\Sigb}{\boldsymbol{\Sigma}}
\newcommand{\Sigbs}{\boldsymbol{\Sigma^*}}
\newcommand{\omegab}{\boldsymbol{\omega}}
\newcommand{\epsb}{\boldsymbol{\epsilon}}
\newcommand{\ub}{\mathbf{u}}
\newcommand{\eb}{\mathbf{e}}\newcommand{\vv}[1]{\underline{#1}}\newcommand{\ev}{\vv{e}}
\newcommand{\rv}{\vv{r}}\newcommand{\TT}[1]{\underline{\underline{#1}}}\newcommand{\omb}{\mathbf{\omega}}
\newcommand{\Ub}{\mathbf{U}}\newcommand{\xb}{\mathbf{x}}\newcommand{\rb}{\mathbf{r}}
\newcommand{\ssb}{\mathbf{s}}\newcommand{\Xb}{\mathbf{X}}\newcommand{\Rey}{\mbox{\textit{Re}}}
\newcommand{\mean}[1]{\left\langle #1\right\rangle}
\newcommand{\ddp}{[p]^\pm}\newcommand{\taub}{\mbox{\boldmath$\tau$}}\newcommand{\Fr}{\mbox{\textit{Fr}}}
\let\grad\nabla\newcommand{\z}{\zeta}\newcommand{\kk}{\kappa}\newcommand{\tkk}{\tilde{\kappa}}
\newcommand{\e}{\varepsilon}\newcommand{\zb}{\bar{\zeta}}\let\grad\nabla\let\bcdot\cdot
\newcommand{\half}{{\textstyle\frac{1}{2}}}
\newcommand{\textfrac}[2]{{\textstyle\frac{#1}{#2}}}
\newcommand{\LF}[1]{{#1}^{\mathrm{LF}}}\newcommand{\Lap}[1]{{#1}^{\mathrm{L}}}
\newcommand{\ds}{*\!*}\newcommand{\cond}[2]{\frac{\mathrm{D} #1}{\mathrm{D} #2}}
\newcommand{\pard}[2]{\frac{\partial #1}{\partial #2}}\newcommand{\totd}[2]{\frac{\mathrm{d}#1}{\mathrm{d}#2}}
\newcommand{\pardd}[3]{\frac{\partial^2 #1}{\partial #2 \partial #3}}
\newcommand{\Real}{\mbox{Re}}\newcommand{\Imag}{\mbox{Im}}
\newcommand{\Fpint}{=\!\!\!\!\!\!\!\int}
\newcommand{\txi}{\tilde\xi}\newcommand{\dxi}{\delta\xi}
\newcommand{\tpsi}{\tilde\psi}\newcommand{\dpsi}{\delta\psi}
\newcommand{\change}[1]{\textcolor{red}{ #1}}
\makeatletter
\def\sgn{\mathop{\operator@font sgn}}
\makeatother

\title{Energy harvesting efficiency of piezoelectric flags in axial flows}
\author{S\'ebastien Michelin}
\email{sebastien.michelin@ladhyx.polytechnique.fr}
\affiliation{LadHyX -- D\'epartement de M\'ecanique, Ecole
  polytechnique, 91128 Palaiseau Cedex, France.}
\author{Olivier Doar\'e}
\email{olivier.doare@ensta-paristech.fr}
\affiliation{ENSTA Paristech, Unit\'e de M\'ecanique, Chemin de la Huni\`ere, 91761 Palaiseau, France}

\date{\today}

\begin{abstract}
Self-sustained oscillations resulting from fluid-solid instabilities, such as the flutter of a flexible flag in axial flow, can be used to harvest energy if one is able to convert the solid energy into electricity. Here, this is achieved using piezoelectric patches attached to the surface of the flag that convert the solid deformation into an electric current powering purely resistive output circuits. Nonlinear numerical simulations in the slender-body limit, based on an explicit description of the coupling between the fluid-solid and electric systems, are used to determine the harvesting efficiency of the system, namely the fraction of the flow kinetic energy flux effectively used to power the output circuit, and its evolution with the system's parameters. The role of the tuning between the characteristic frequencies of the fluid-solid and electric systems is emphasized, as well as the critical impact of the piezoelectric coupling intensity. High fluid loading, classically associated with destabilization by damping, leads to greater energy harvesting, but with a weaker robustness to flow velocity fluctuations due to the sensitivity of the flapping mode selection. This suggests that a control of this mode selection by a careful design of the output circuit could provide some opportunities of improvement for the efficiency and robustness of the energy harvesting process. 
\end{abstract}
\maketitle

\section{Introduction}
The limited availability and environmental impact of fossile fuels motivate the development of renewable energy sources. Significant research efforts are currently made to propose energy harvesting concepts and prototypes converting the kinetic energy of geophysical flows such as winds, rivers and oceanic or tidal currents into electricity \citep{westwood2004}. In parallel, a particular attention is currently given to systems able to produce limited amount of energy from different vibration sources in order to power remote or isolated devices \citep*{sodano2004}. Classical fluid-solid couplings and instabilities such as vortex-induced vibrations, galloping and flutter in axial flows effectively act as energy extraction mechanisms as they enable an energy transfer from the incoming flow to the solid body, and can therefore be used to produce electricity using displacement-based (e.g. electromagnetic converters) or deformation-based (e.g. piezoelectric materials) conversion mechanisms \citep{bernitsas2008,barrero2010,zhu2009c,singh2012b} . Because they are based on fundamentally-different mechanisms, such flow energy harvesters may be attractive complements to the existing wind- and water-turbines technologies, and properly assessing fundamental upper bounds on their respective efficiency is therefore of critical importance.

A flexible plate placed in an axial flow becomes unstable to flutter above a critical flow velocity when the destabilizing pressure forces dominate the stabilizing effect of the structure's rigidity \citep*{kornecki1976,paidoussis2004,shelley2011}. This critical velocity depends on the plate's properties (e.g. density, size and rigidity) and can therefore be adjusted in the system's design to be lower than the typical flow velocity. This so-called flapping flag instability leads to self-sustained large-amplitude flapping of the plate in the form of traveling bending waves \citep{connell2007,eloy2008,alben2008,michelin2008}, that can be used to produce electricity using, for example, piezoelectric patches attached to the plate's surface \citep{allen2001,dunnmon2011,giacomello2011,akcabay2012}. 

An important research effort is required in order to assess the amount of energy that can be harvested using such devices and investigate possible intrinsic limits or potential optimization strategies of their efficiency. In a theoretical or numerical framework, the conversion mechanism and output circuit must be described, to properly include the coupling of the fluid-solid and electric systems. Energy harvesting eventually amounts to an extraction of energy from the solid dynamics. Hence, a first and simpler model for the harvesting mechanism is  an additional structural damping (e.g. Kelvin-Voigt), and assessing the system's efficiency is then equivalent to determining how much energy can be dissipated by the flapping structure \citep{tang2009b,zhu2009c,singh2012b}. Indeed, increasing damping would lead to a larger energy dissipation but eventually will re-stabilize the system and reduce its harvesting efficiency. Although simple to implement, this representation is not complete as it assumes that energy is instantaneously and immediately dissipated and can not represent the dynamics of the electrical circuit or of the coupling mechanism.

The originality of the present work is to propose instead a fully-coupled description of a fluid-solid-electric system, namely a flexible plate in axial flow covered with piezoelectric patch pairs powering simple resistive elements. Recently, \citet{doare2011} followed this approach to study the impact of the piezoelectric coupling on the linear stability of a two-dimensional plate and on the solid-electric energy transfers. In particular, the role of the tuning of the fluid-solid and electric characteristic time-scales was emphasized, and a destabilization by the piezoelectric coupling was identified in the case of large fluid loading, associated with the destabilization by damping of negative energy waves \citep{benjamin1963,doare2010}. The present study extends this approach to study numerically the nonlinear dynamics of this fluid-solid-electric system in the case of a slender flexible plate, and to determine its harvesting efficiency. Here, the system's efficiency is defined following the classical definition used for wind-turbines, as the ratio of the mean power output and of the mean kinetic energy flux through the section occupied by the device in the flow. In that sense, it differs from the measures of efficiency used in other existing studies  \citep{tang2009b,dunnmon2011}. 

In \S\ref{sec:model}, the model used to describe the dynamics of the flapping piezoelectric flag is presented. Section~\ref{sec:linear} presents a short summary of the linear stability results in the case of a slender plate. In \S\ref{sec:nl}, the numerical solution of the coupled dynamics is addressed, the system's efficiency is defined and the impact of the different system parameters on this efficiency is discussed. Finally, conclusions and perspectives are presented in \S\ref{sec:conclusions}.


\section{Presentation of the fully-coupled model}
\label{sec:model}
\subsection{Piezoelectric flag dynamics}
\begin{figure}
\begin{center}
\includegraphics[width=.9\textwidth]{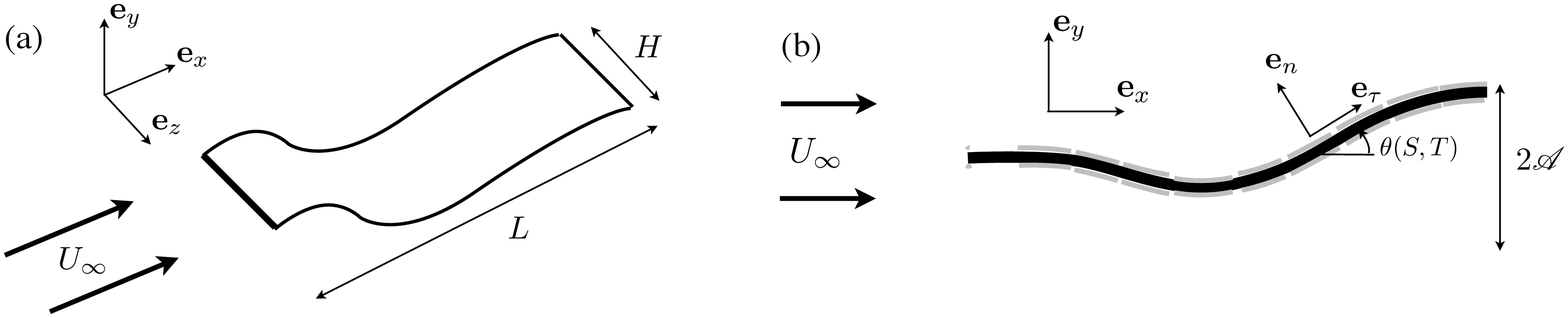}
\caption{(a) Slender flexible plate flapping in a uniform axial flow. (b) Two-dimensional flapping of a slender flexible plate covered with pairs of piezoelectric patches.}\label{fig:piezoflag_schema}
\end{center}
\end{figure}

The system considered here consists of a rectangular flexible plate of length $L$, width $H$ and thickness $h$ ($h\ll H, L$) placed in a steady flow of density $\rho$ and velocity $U_\infty$. The plate is inextensible and clamped at its leading edge; for simplicity, only purely planar motions of the plate are considered, so that the plate's position $\Xb$ is only a function of the streamwise curvilinear coordinate $S$ and time $T$, and the solid does not experience any spanwise displacement nor twist. The local orientation of the flag with respect to the horizontal axis is noted $\theta(S,T)$ (Figure \ref{fig:piezoflag_schema}). In the following, lineic quantities will be defined per unit length in $S$.

The surface of the plate is covered by pairs of piezoelectric patches (Figure~\ref{fig:piezoflag_schema}b) with streamwise length $l\ll L$ and width $H$. The negative electrodes of each patch are shunted through the plate and the positive electrodes are connected to the output circuit. The deformation of the flag is coupled to the output circuit through the piezoelectric coupling: (i) stretching and compression of the patches due to the local curvature induces charge transfers between each patch's electrodes and (ii) an electric voltage applied to its electrodes results in an additional internal torque on the piezoelectric patch and on the flag. Considering the limit of a continuous coverage by patches of infinitesimal length \citep*{bisegna2006,doare2011}, the local electric state can be described in terms of the electric voltage between the positive electrodes of each patch, $V(S,T)$, and the charge transfer $Q(S,T)$ per unit length in the streamwise direction. In this limit, which differs from the single-patch approach of \citet{akcabay2012}, both quantities are continuous functions of $S$ and $T$, and the piezoelectric coupling imposes that:
\begin{align}
Q&=cV+\chi^*\pard{\theta}{S},\label{eq:piezo1}\\
\mathcal{M}&=B\pard{\theta}{S}-\chi^* V,\label{eq:piezo2}
\end{align}
where $\mathcal{M}$ is the total internal torque in the piezoelectric flag, and $c$ and $\chi^*$ are the lineic capacitance and piezoelectric coupling coefficient, directly related to the material and geometric properties of the patch pair \citep{doare2011}. An Euler--Bernoulli model is assumed for the dynamics of the piezoelectric flag with $B$ the effective flexural rigidity of the three-layer piezoelectric plate \citep[for more details, see][]{lee1989,doare2011}.

\begin{figure}
\begin{center}
\includegraphics[width=.85\textwidth]{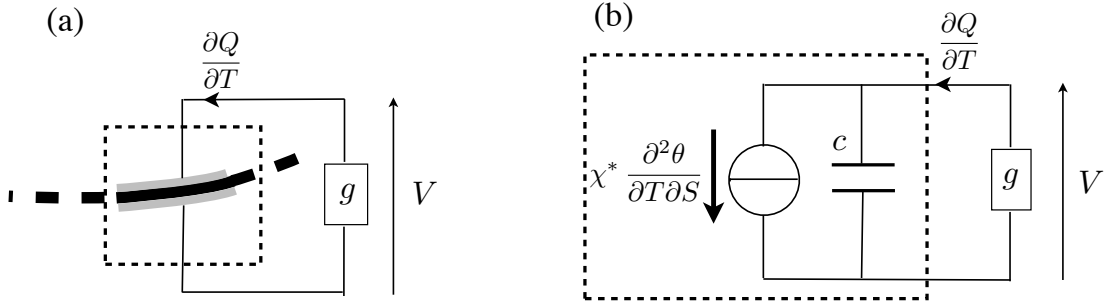}
\caption{(a) Piezoelectric patch pair powering a purely resistive circuit. (b) The piezoelectric patch pair is characterized by the current generated by its deformation and by its capacitance $c$.}\label{fig:piezoflag_schema2}
\end{center}
\end{figure}

The positive electrodes are connected to a purely resistive circuit of lineic conductivity $g$ (Figure \ref{fig:piezoflag_schema2}a), such that
\begin{equation}
\pard{Q}{T}+gV=0.
\end{equation}


The conservation of momentum and inextensibility condition for the flag lead to: 
\begin{align}
\mu\pard{^2\Xb}{T^2}&=\pard{}{S}\left[F_T\mathbf{e}_\tau-\pard{\mathcal{M}}{S}\mathbf{e}_n\right]+\Fb_\textrm{fluid},\label{eq:beam}\\
\pard{\Xb}{S}&=\mathbf{e}_\tau,\label{eq:inext}
\end{align}
with $\mu$ the lineic mass of the piezoelectric flag, $F_T(S,T)$ the local tension, acting as a Lagrangian multiplier to enforce the plate's inextensibility \eqref{eq:inext}, and $\mathcal{M}$ the internal piezo-elastic torque in \eqref{eq:piezo2}. The following clamped-free boundary conditions must also be satisfied:
\begin{align}
\Xb=0,\qquad \theta=0\qquad \textrm{at    } S=0,\\
\mathcal{M}=\pard{\mathcal{M}}{S}=F_T=0\qquad \textrm{at   }S=L.
\end{align}

The conservation of mechanical and electrical energy takes the following form:
\begin{equation}
\totd{}{T}\left(\mathscr{E}_k+\mathscr{E}_{p}\right)=\mathscr{W}_p-\mathscr{F},\qquad \totd{\mathscr{E}_{el}}{T}=\mathscr{F}-\mathscr{P},\label{eq:energy}
\end{equation}
where
\begin{equation}
\mathscr{E}_k=\int_0^L\frac{1}{2}\mu\left|\pard{\Xb}{T}\right|^2\dd S,\qquad \mathscr{E}_p=\int_0^L\frac{1}{2}B\left(\pard{\theta}{S}\right)^2\dd S,\qquad \mathscr{E}_{el}=\int_0^L\frac{1}{2}cV^2\dd S
\end{equation}
are respectively the kinetic and potential elastic energy of the flag, and the energy stored in the capacitance of the piezoelectric elements, and
\begin{equation}
\mathscr{W}_p=\int_0^L\Fb_\textrm{fluid}\cdot\pard{\Xb}{T}\dd S,\qquad \mathscr{F}=-\chi^*\int_0^LV\pardd{\theta}{T}{S}\dd s,\qquad\mathscr{P}=-\int_0^L V\pard{Q}{T}\dd S,
\end{equation}
are the rate of work of the fluid forces, the rate of energy transfer from the solid to the electric circuit and the power used in the output circuit, respectively. For a purely resistive circuit $\mathscr{P}$ is always strictly positive, and in permanent periodic regime, $\mean{\mathscr{P}}=\mean{\mathscr{F}}=\mean{\mathscr{W}_p}$, with $\mean{\cdot}$ the time-averaging operator.

\subsection{Fluid modeling - Lighthill's theory}
The relative motion of the solid body with respect to the incoming flow results in fluid forces $\Fb_\textrm{fluid}$ applied on its surface. In the particular limit of a slender body ($H\ll L$) and for a purely potential flow, the extension of Lighthill's Elongated Body Theory to large amplitude displacements leads to the following leading order expression for the reactive fluid forces $\Fb_\textrm{reac}$ associated with the local transverse motion of each cross section along the plate:
\begin{equation}\label{eq:lighthill}
\Fb_\textrm{reac}=-m_a\rho H^2\left(\pard{U_n}{T}-\pard{}{S}\left(U_nU_\tau\right)+\frac{1}{2}U_n^2\pard{\theta}{S}\right)\eb_n.
\end{equation}
with $m_a$ the non-dimensional added mass coefficient of the local cross-section, namely $m_a=\pi/4$ for a flat plate. In \eqref{eq:lighthill}, $U_\tau$ and $U_n$ are respectively the tangential and normal components of the local relative velocity of the solid with respect to the incoming flow:
\begin{equation}
\Ub=\pard{\Xb}{T}-U_\infty\eb_x=U_\tau\eb_\tau+U_n\eb_n.
\end{equation}
Initially proposed by \citet{lighthill1971}, this so-called Large Amplitude Elongated Body Theory (LAEBT) was recently shown to provide a good estimate of the transverse fluid forces, in comparison with RANS simulations on a towed and deforming fish body \citep*{candelier2011}. However, \citet{candelier2011} emphasized that this purely reactive formulation can not by itself represent properly the deformation amplitude of freely-moving bodies, as such effects as drag and separation will be significant and must be accounted for by an additional resistive component $\Fb_\textrm{resist}$ \citep[see for example][]{taylor1952}. In the case of a freely-flapping slender body, \citet*{singh2012} indeed observed that the purely reactive model would lead to non-physical overestimates of the flapping amplitude. Following \citet{eloy2012} and \citet{singh2012b}, the present model only retains the resistive drag associated with the plate's normal displacement 
\begin{equation}\label{eq:resistive}
\Fb_\textrm{resist}=-\frac{1}{2}\rho H C_D U_n\left|U_n\right|\eb_n,
\end{equation}
with $C_D=1.8$ for a flat plate in transverse flows. 

The reactive part of the LAEBT corresponds to the asymptotic limit of the potential flow equations when $H/L\ll 1$ \citep{candelier2011}, but the recent work of \citet{eloy2012} showed nonetheless, using comparisons with wind-tunnel experiments, that the combination of the reactive and resistive components \eqref{eq:lighthill} and \eqref{eq:resistive} can provide a good prediction of the flapping properties of the plate even when $H/L=O(1)$. In the following, an aspect ratio $H^*=H/L=0.5$ will therefore be considered.

\subsection{Non-dimensional equations}
Equations \eqref{eq:piezo1}--\eqref{eq:resistive} are non-dimensionalized using $L$, $L/U_\infty$, $\rho H L^2$, $U_\infty\sqrt{\mu/c}$ and $U_\infty\sqrt{\mu\,c}$ as characteristic length, time, mass, voltage and charge density, respectively:
\begin{align}
\pard{^2\xb}{t^2}&=\pard{}{s}\left[f_T\eb_\tau -\pard{}{s}\left(\frac{1}{U^{*2}}\pard{\theta}{s}-\frac{\alpha}{U^*}v\right)\eb_n\right]+M^*f_\textrm{fluid} \,\eb_n,\label{eq:beamnd}\\
f_\textrm{fluid}&=-\frac{1}{2}C_d u_n\,|u_n|-m_a H^*\left(\pard{u_n}{t}-\pard{}{s}(u_nu_\tau)+\frac{1}{2}u_n^2\pard{\theta}{s}\right),\\
q&=v+\frac{\alpha}{U^*}\pard{\theta}{s},\label{eq:piezond}\\
\beta\pard{q}{t}&+v=0.\label{eq:ohmnd}
\end{align}
and the tension $f_T$ is obtained using the inextensibility condition \citep[see for example][]{michelin2008,alben2009}
\begin{equation}
\pard{\xb}{s}=\mathbf{e}_\tau.
\end{equation}
The clamped-free boundary conditions become
\begin{align}
\textrm{at   }s=0,&\quad\xb=0,\quad \theta=0\\
\textrm{at   }s=1,&\quad f_T=\pard{\theta}{s}-\alpha U^*v=\pard{^2\theta}{s^2}-\alpha U^* \pard{v}{s}=0.\label{eq:bcnd2}
\end{align}
Five non-dimensional parameters characterize the system, namely the fluid-solid inertia ratio, the non-dimensional velocity $U^*$, the coupling coefficient $\alpha$, the tuning coefficient of the fluid-solid and electric system $\beta$ and the aspect ratio of the plate $H^*$:
\begin{equation}
M^*=\frac{\rho H L}{\mu},\quad U^*=U_\infty L\sqrt{\frac{\mu}{B}},\quad \alpha=\frac{\chi^*}{\sqrt{B c}},\quad \beta=\frac{c\,U_\infty}{gL},\quad H^*=\frac{H}{L}\cdot \label{eq:nondimparam}
\end{equation}
The originality of the present work is to offer a full description of the fluid-solid-electric system. Equations~\eqref{eq:beamnd}, \eqref{eq:piezond} and \eqref{eq:ohmnd} show that the effect of the piezoelectric coupling is more complex than the simple Kelvin--Voigt damping model generally assumed for simplicity in most studies on energy harvesting flags \citep{tang2009b,singh2012}. Indeed, combining \eqref{eq:piezond} and \eqref{eq:ohmnd}, one obtains
\begin{equation}\label{eq:piezo_damp}
\beta\dot{v}+v=-\frac{\alpha\beta}{U^*}\pardd{\theta}{s}{t}.
\end{equation}
Equation \eqref{eq:piezo_damp} shows that the effective damping introduced by the piezoelectric is frequency-dependent. In fact, a Kelvin--Voigt damping model could only be recovered in the particular limit of $\beta\ll 1$ and finite $\alpha\beta/U^*$. However, this asymptotic limit is unlikely to be achieved in practice because of the material restrictions on the coupling coefficient $\alpha$ for currently-available piezoelectric materials \citep{doare2011}. 

\section{Linear stability analysis}
\label{sec:linear}
The linear stability of the piezoelectric flag is first analyzed to identify the impact of the piezoelectric coupling and output circuit on the stability properties of the system, and also identify the operating regime of the harvesting devices, namely the parameter values for which self-sustained oscillations can develop. The present linear study only differs from  that in \citet{doare2011} by the fluid model considered, that corresponds to a different range for the plate's aspect ratio, therefore only the main results will be reminded and the reader is referred to this previous contribution for more in-depth analysis of the linear stability.

\begin{figure}
\begin{center}
\begin{tabular}{cc}
\includegraphics[width=.45\textwidth]{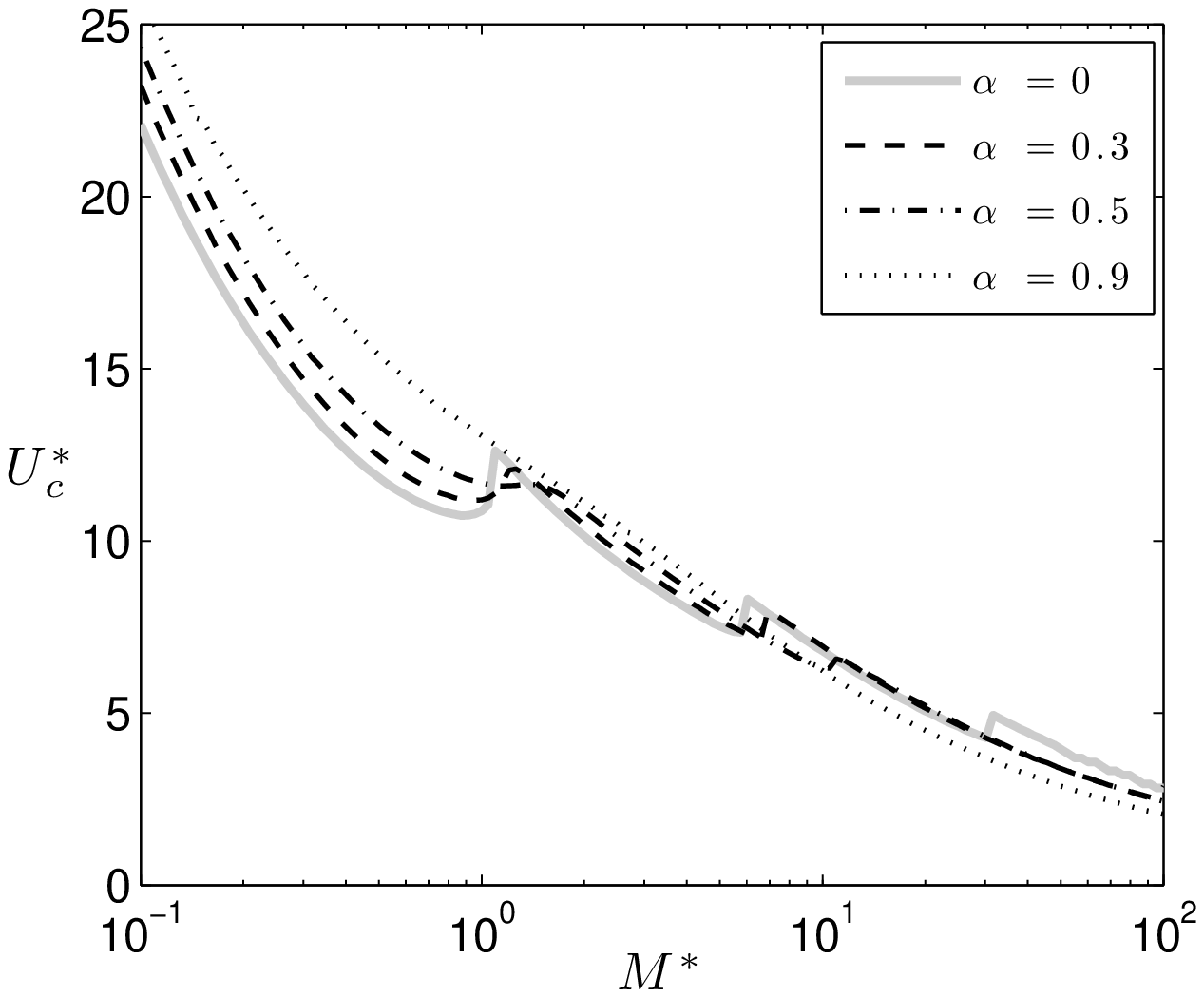} &
\includegraphics[width=.45\textwidth]{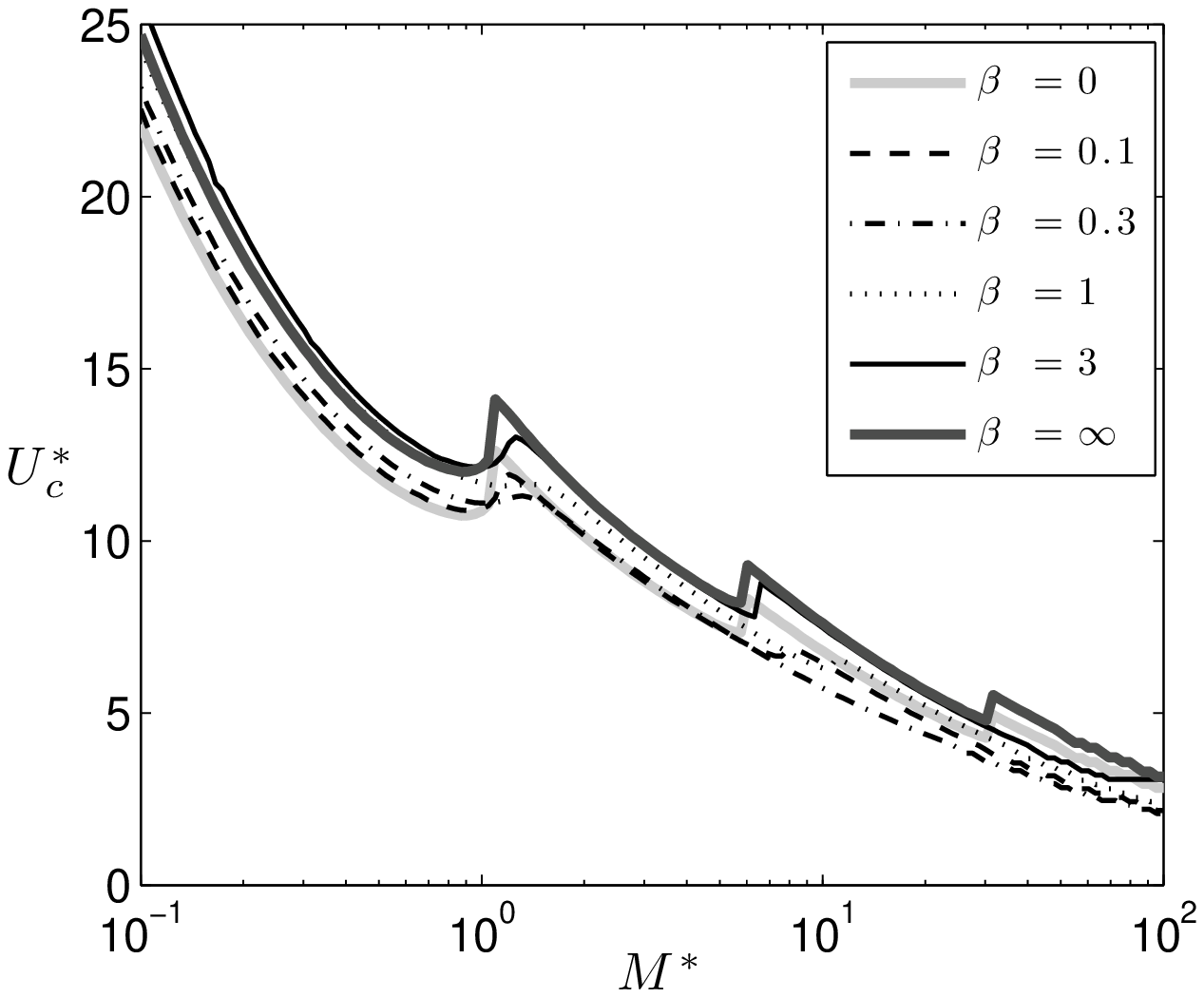}
\end{tabular}
\caption{(Left) Critical velocity threshold $U^*_c$ as a function of the mass ratio $M^*$ for $\beta=1$ and varying $\alpha$. (Right) Critical velocity threshold $U^*_c$ as a function of the mass ratio $M^*$ for $\alpha=0.5$ and varying $\beta$.}\label{fig:threshold}
\end{center}
\end{figure}

The displacement of the flag is purely vertical and noted $y(s,t)\ll 1$. At leading order, \eqref{eq:beamnd}--\eqref{eq:bcnd2} simplify into the following linear systems for $(y,v)$:
\begin{align}
(1+m_aM^*H^*)\pard{^2y}{t^2}&+2m_aM^*H^*\pardd{y}{t}{s}+m_aM^*H^*\pard{^2y}{s^2}+\frac{1}{U^{*2}}\pard{^4y}{s^4}-\frac{\alpha}{U^*}\pard{^2v}{s^2}=0,\label{eq:beamlin}\\
&\beta\pard{v}{t}+v+\frac{\alpha\beta}{U^*}\frac{\partial^3y}{\partial s^2\partial t}=0,
\end{align}
with boundary conditions:
\begin{align}
\textrm{at   }s=0,&\quad y=\pard{y}{s}=0\\
\textrm{at   }s=1,&\quad\pard{^2y}{s^2}-\alpha U^*v=\pard{^3y}{s^3}-\alpha U^* \pard{v}{s}=0.\label{eq:bclin2}
\end{align}
Searching for solutions of the form $[y,v]=\Real\left([\tilde{Y},\tilde{V}]\ee^{-\ci\omega t}\right)$, \eqref{eq:beamlin}--\eqref{eq:bclin2} become an eigenvalue problem for $\omega$ and $[\tilde{Y}(s),\tilde{V}(s)]$, that is solved numerically using a Chebyshev collocation method to determine the stability of the piezoelectric flag, and in particular the critical velocity above which the flag becomes unstable (Figure \ref{fig:threshold}).

The piezoelectric coupling $\alpha$ enables the transfer of energy from the fluid-solid system to the electrical circuit where part of it is dissipated, resulting in a net damping on the solid motion. This additional damping is therefore expected to increase the critical velocity in comparison with the uncoupled flag ($\alpha=0$), an effect indeed observed for $M^*\lesssim 1$ (Figure \ref{fig:threshold}a). At larger $M^*$, the piezoelectric coupling instead destabilizes the system, at least initially. This destabilization by damping was previously reported in the case of a two-dimensional flag by \citet{doare2011}, and is associated with the existence of negative energy waves in the local stability analysis of the non-dissipative flag \citep{benjamin1963}. From an energy harvesting point of view, it increases the operating range of the piezoelectric flag as self-sustained oscillations develop for lower velocities.

For a fixed piezoelectric coupling, $\beta$ measures the tuning of the fluid-solid and electric time-scales of the system. When forced by the flag at a frequency much lower than $1/\tau_{RC}=g/c$ ($\beta\ll 1$), the output resistive elements are seen by the piezoelectric as short circuits, and the voltage at the electrodes remains negligible. The critical velocity is therefore equal to that of the uncoupled piezoelectric flag ($\alpha=0$) as no piezoelectric feedback is applied on the structure. For a large forcing frequency ($\beta\gg 1$) however, the resistive elements are seen as open circuits, and from Eq.~\eqref{eq:piezo1}, the voltage at the piezoelectric's electrodes is proportional and opposite to the curvature: the piezoelectric coverage then acts as an additional rigidity on the system. Between these two limit regimes, a destabilization is observed for large $M^*$ which corresponds to the destabilization mechanism mentioned above (Figure \ref{fig:threshold}b).

These results confirm and extend to the slender-body limit the conclusions of the infinite-span flag analysis of \citet{doare2011}. It is worth noting that the results obtained with both models differ mostly at low $M^*$, consistently with the results of \citet{eloy2007} on the impact of aspect ratio on the flag stability.

\section{Non-linear dynamics of a piezoelectric flag}
\label{sec:nl}
To determine the amount of energy that can be produced using such a system, the nonlinear dynamics of the piezoelectric flag must be studied, in particular to determine its flapping amplitude and frequency.

\subsection{Non-linear simulations and energy harvesting efficiency}
Following \citet{alben2009}, the nonlinear system \eqref{eq:beamnd}--\eqref{eq:bcnd2} is integrated numerically in time using a second-order accurate implicit method, and spatial derivatives are computed using Chebyshev collocation. Starting from rest ($\theta(s,t<0)=0$), the flag is excited by a small perturbation in the vertical component of the upstream flow. The harvested energy is computed as the temporal average of the non-dimensional power $\mathcal{P}=\mathscr{P}/(\rho U_\infty^3HL)$ dissipated in the resistive elements in permanent regime:
\begin{equation}
\mathcal{Q}=\mean{\mathcal{P}}=\mean{\frac{1}{\beta M^*}\int_0^1v^2\dd s}\cdot
\end{equation}
In the previous equation, the temporal average is understood and computed as follows: when the system converges to limit-cycle oscillations, it is defined as the mean value over a period of oscillation, but when no limit-cycle oscillation can be identified, it is computed as the statistical average of $\mathscr{P}$ over a long enough time frame. Similarly, the non-dimensional flapping amplitude $\mathcal{A}=\mathscr{A}/L$ is defined from the trailing edge displacement $y_e(t)$ as a measure of the peak flapping amplitude:
\begin{equation}
\mathcal{A}=\sqrt{2\mean{y_e^2}}.
\end{equation}
The harvesting efficiency of the system, $\eta$, is defined as the fraction of the fluid kinetic energy flux through the cross-section $2\mathscr{A}H$ occupied by the flag (Figure~\ref{fig:piezoflag_schema}) actually transferred to the output circuit, namely
\begin{equation}
\eta=\frac{\mean{\mathscr{P}}}{\frac{1}{2}\rho U_\infty^3\times 2\mathscr{A}H}=\frac{\mathcal{Q}}{\mathcal{A}}\cdot
\end{equation}

\subsection{Non-linear flapping dynamics}
\label{sec:nlflap}
Above the critical velocity $U_c^*$, defined using the linear stability analysis of Section~\ref{sec:linear}, an initial perturbation of the flag's state of rest leads to an exponential growth of the flapping amplitude until saturation is reached, and the permanent regime takes one of the two following forms: (i) a strongly periodic regime characterized by the identification of a limit-cycle in phase-space or (ii) a more complex non-linear regime where no clear limit-cycle can be identified. This transition from periodic to non-periodic regime has been observed in numerous experimental \citep{eloy2012} and numerical studies \citep{connell2007,michelin2008,alben2008}, and has been conjectured to result from the non-linear interactions of different fundamental modes. Limit-cycle oscillations are particularly interesting from an energy harvesting point of view as it provides  steady output current amplitude and frequency.

\begin{figure}
\begin{center}
\includegraphics[width=.85\textwidth]{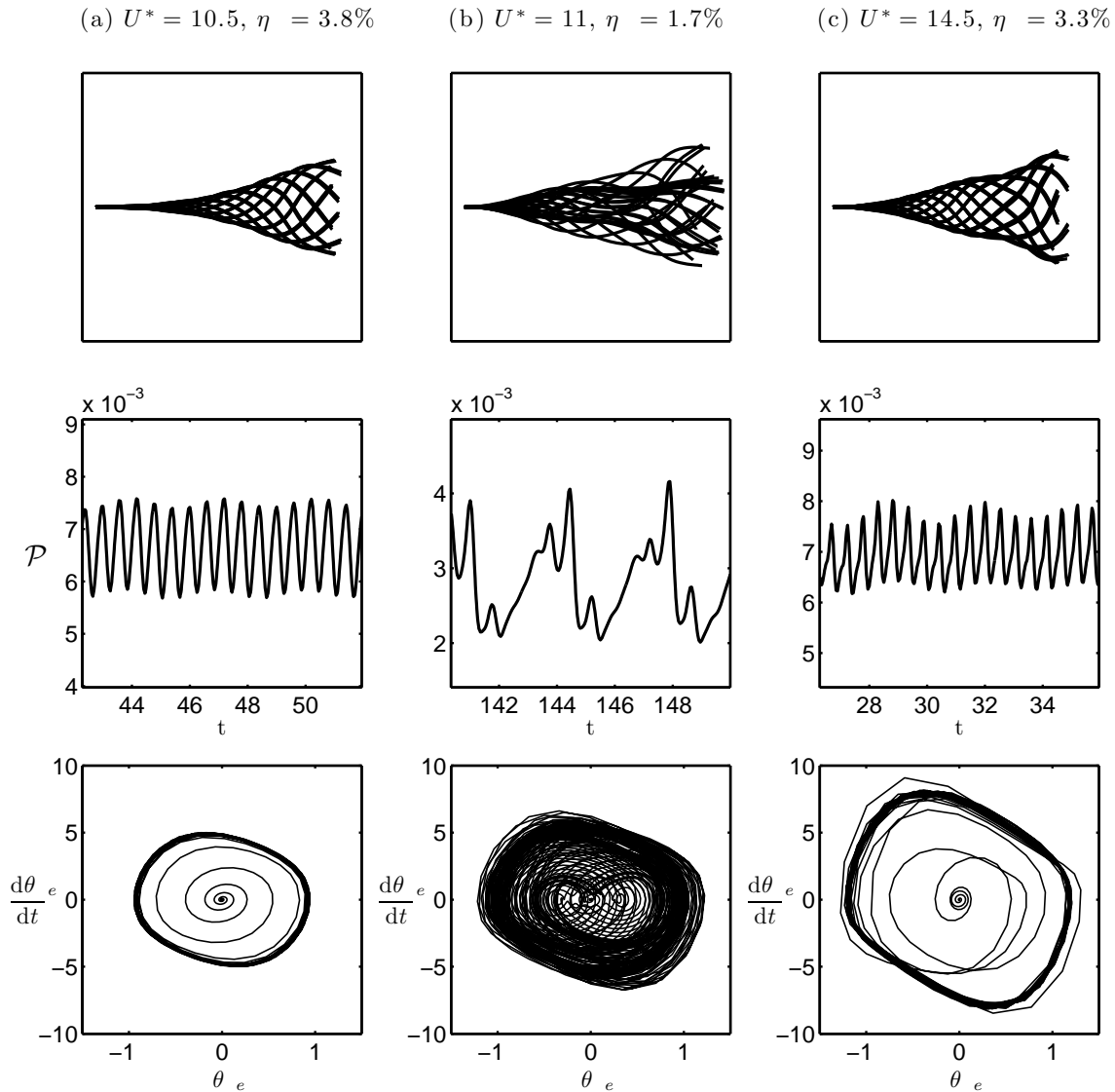}
\caption{Limit-cycle oscillations: (Top) Flapping mode shape, (Center) Time-series of the non-dimensional harvested power $\mathcal{P}(t)$ and (Bottom) Phase-space trajectory for the trailing edge orientation $\theta_e(t)$ for $M^*=10$, $\alpha=0.5$, $\beta=0.158$ and (a) $U^*=10.5$, (b) $11$ and (c) $14.5$ from left to right.}\label{fig:LCO}
\end{center}
\end{figure}

Even below the transition to chaotic flapping, non-periodic flapping regimes can be observed as the systems switches from one flapping mode to another when one of the parameters (e.g. $U^*$) is modified. This mode switch results in a change of flapping amplitude and frequency, but also of the flag kinematics resulting in a modification of the forcing distribution on the piezoelectric elements (Figure \ref{fig:LCO}). 

\subsection{Variations of the energy harvesting efficiency}
In the following, the impact of the different parameters on the harvesting efficiency is presented.
\subsubsection{Effect of the tuning ratio}
\begin{figure}
\begin{center}
\includegraphics[width=.85\textwidth]{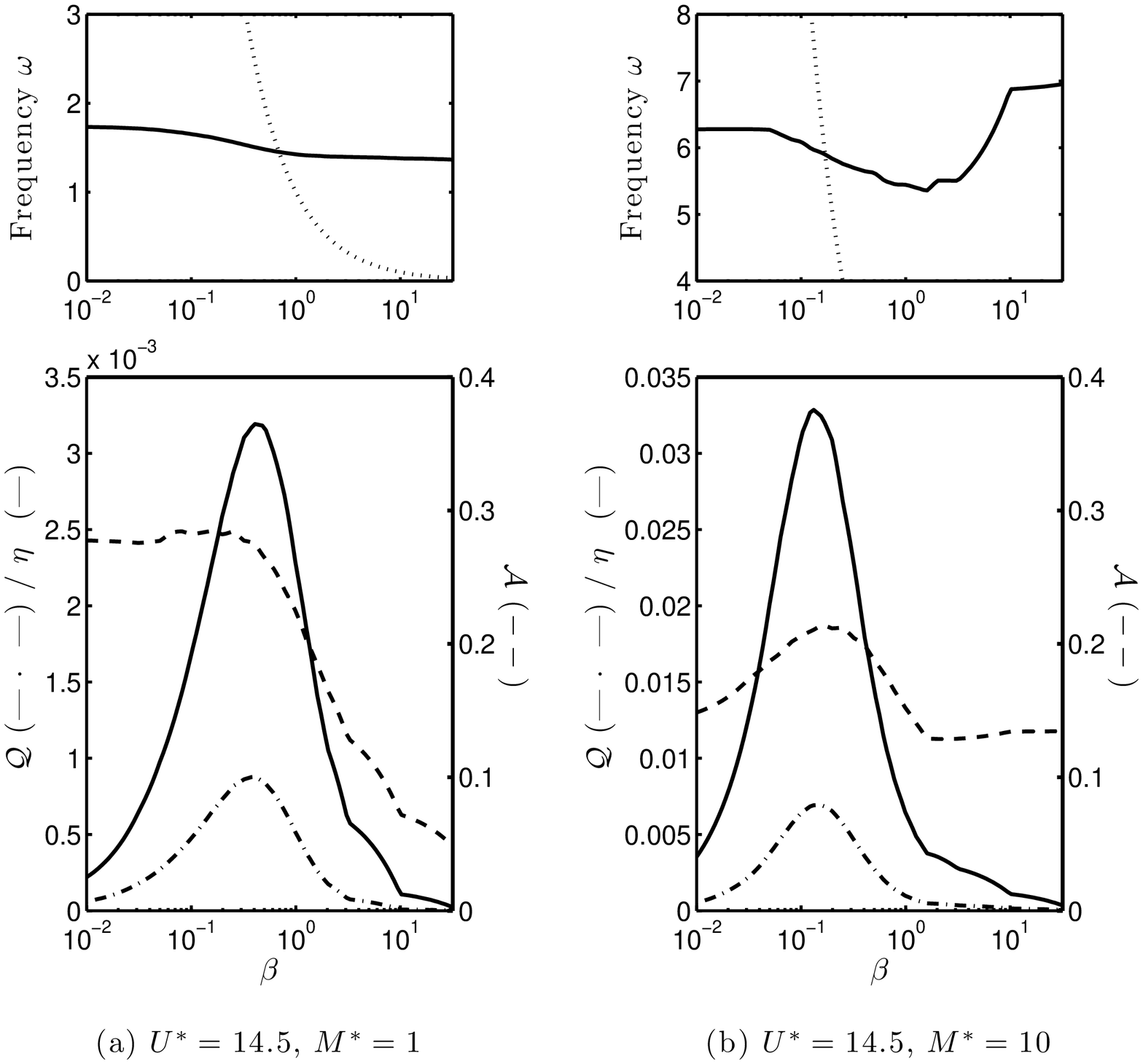}
\caption{Evolution with the tuning ratio $\beta$ of (Top) the non-dimensional flapping frequency $\omega$ and (Bottom) the non-dimensional flapping amplitude $\mathcal{A}$ (dashed), harvested energy $\mathcal{Q}$ (dash-dotted) and harvesting efficiency $\eta$ (solid) for $\alpha=0.5$, $U^*=14.5$ and (a) $M^*=1$ or (b) $M^*=10$. On the top, the dotted line correspond to the variations of $1/\beta$.}\label{fig:AQeta}
\end{center}
\end{figure}

The ratio $\beta=cU_\infty/(gL)$ measures the relative tuning of the fluid-solid and electric time scales,  $\tau_\textrm{adv}=L/U_\infty$ and $\tau_{RC}=c/g$, respectively. All other parameters being fixed, it is observed that the harvested energy efficiency reaches a maximum when $\beta\omega=O(1)$ where $\omega$ is the non-dimensional flapping frequency of the flag (Figure \ref{fig:AQeta}).

The existence of this maximum comes as no surprise: when $\beta\ll 1$ and $\beta\gg 1$, the resistive element acts as a short-circuit or open-circuit, respectively. In both cases, no energy is dissipated and $\eta=0$. $\beta\omega=O(1)$ corresponds to a forcing of the RC output circuit at its characteristic time-scale, which is expected to result in maximum energy dissipation in the resistive element. The forcing frequency is however not a property of the fluid-solid system only, but is instead the result of the nonlinear coupling between the fluid-solid system and the electric output through the piezoelectric material (Figure \ref{fig:AQeta}). Similarly, the flapping amplitude is significantly modified when $\beta$ is varied: in particular, for $M^*=1$ (Figure \ref{fig:AQeta}a), a sharp drop in the flapping amplitude is observed as $\beta$ is increased. 

This result is also confirmed on Figures~\ref{fig:eff_scanUbeta} and \ref{fig:freq}. For each value of $U^*$, an optimal tuning ratio can be determined and the optimal $\beta$ is a decreasing function of $U^*$ (Figure \ref{fig:eff_scanUbeta}b). This is consistent with the observed increase in flapping frequency $\omega$ with $U^*$ (Figure~\ref{fig:freq}) and the criterion $\beta\omega=O(1)$ for optimal energy harvesting.

\subsubsection{Effect of the flow velocity}
\begin{figure}
\begin{center}
\includegraphics[width=.9\textwidth]{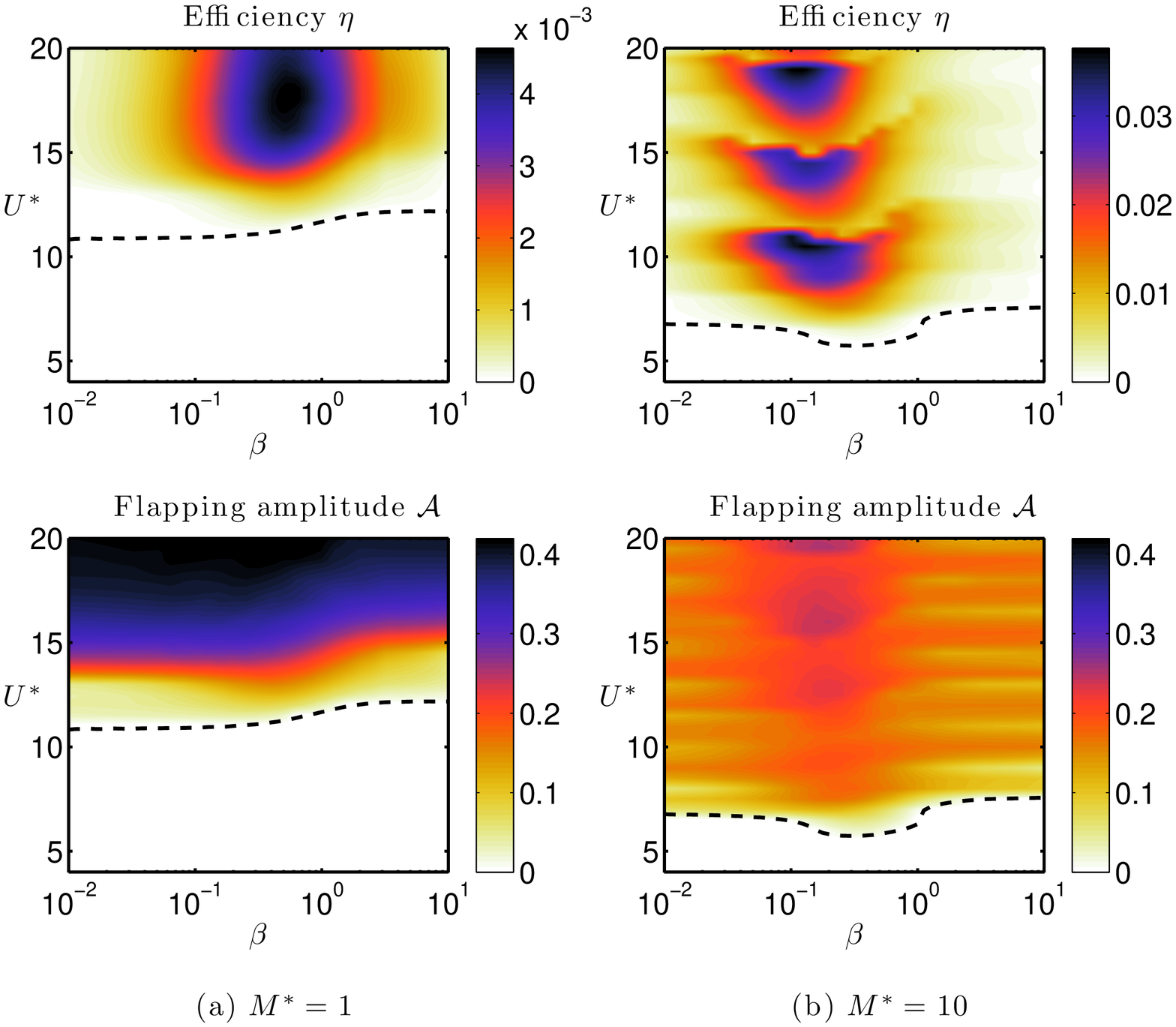}
\caption{(Top) Harvesting efficiency $\eta$ and (Bottom) flapping amplitude $\mathcal{A}$ as a function of the tuning ratio $\beta$ and the non-dimensional flow velocity $U^*$ for (a) $M^*=1$ and (b) $M^*=10$. For both cases, $\alpha=0.5$. The black dashed line corresponds to the instability threshold $U_c^*$ below which $\eta=\mathcal{A}=0$.}\label{fig:eff_scanUbeta}
\end{center}
\end{figure}

\begin{figure}
\begin{center}
\begin{tabular}{cc}
\subfigure[$\beta=0.03$]{\includegraphics[width=.45\textwidth]{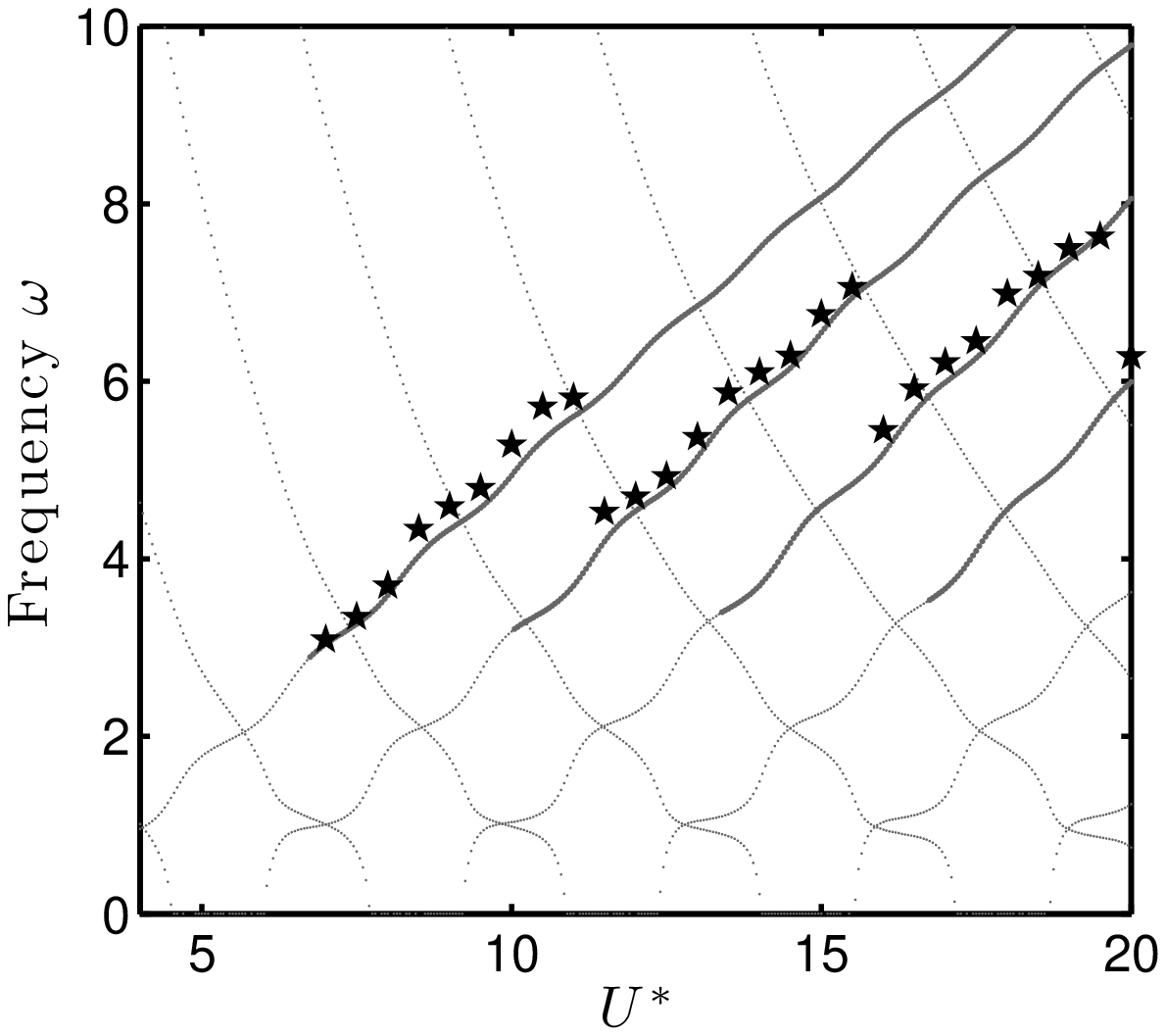}} &
\subfigure[$\beta=0.31$]{\includegraphics[width=.45\textwidth]{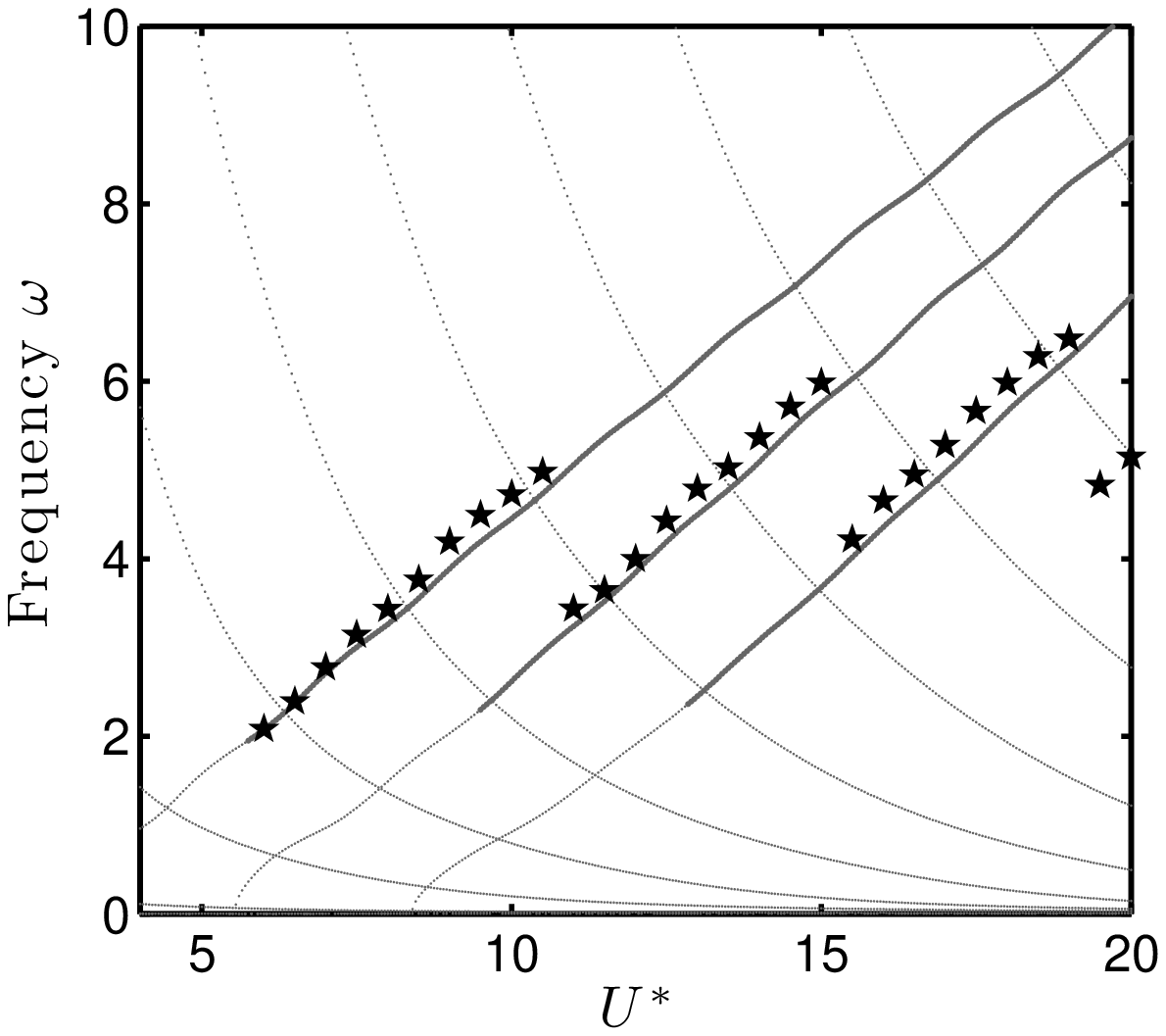}} 
\end{tabular}
\caption{Evolution with $U^*$ of the limit-cycle dominant frequency $\omega$ (black star) for $M^*=10$, $\alpha=0.5$ and (a) $\beta=0.03$ and (b) $\beta=0.31$. In each case, the frequencies of the different linear modes are shown: light dotted lines correspond to stable modes and thick grey lines to unstable modes. On each figure, from left to right, unstable frequencies correspond to flapping modes of increasing order and decreasing characteristic wavelength.}\label{fig:freq}
\end{center}
\end{figure}

Previous experimental results on the dynamics of flexible flags have established that the flapping amplitude is in general an increasing function of the non-dimensional velocity $U^*$ above the instability threshold \citep{shelley2005,eloy2008,eloy2012}, before saturation of this flapping amplitude is reached. For a given flapping mode shape and frequency, the harvested power $\mathcal{P}$ varies quadratically with the amplitude $\mathcal{A}$, therefore it is expected that raising $U^*$ will lead to an increase in the system's efficiency. This is confirmed partially in Figure \ref{fig:eff_scanUbeta}: when a given flapping mode remains dominant, $\eta$ is indeed an increasing function of $U^*$, mainly due to the associated increase in flapping amplitude. However, when a mode switching event occurs as described in Section~\ref{sec:nlflap}, a sudden decrease of the efficiency is observed, mainly associated with a reduction in the flapping frequency (Figure~\ref{fig:freq}).

Figure \ref{fig:freq} shows that the nonlinear flapping frequency is very close to the frequency of one of the unstable linear modes of the piezoelectric flag. A mode switching event, as $U^*$ is increased, consists of a transition from one linearly unstable mode to another with lower frequency. A study of the associated linear growth rate however does not show any coincidence of such event with a change in the most unstable linear mode, and this mode switching event is therefore the result of a purely nonlinear mechanism. Figure \ref{fig:eff_scanUbeta} shows that such mode switching events take place at lower values of $U^*$ for lighter flags (large $M^*$) while for $M^*\lesssim 1$, no such even is detected below $U^*=20$.

This mode selection mechanism is also observed for a flapping flag without any piezoelectric ($\alpha=0$). Regardless of its origin, its importance is however essential for the performance of the energy harvester: as long as the same nonlinear flapping mode can be maintained, the efficiency of the system increases with $U^*$ and the occurrence of a mode switching event results in an important performance loss for the device. A better understanding of this phenomenon and, in particular, of the impact of the piezoelectric coupling on the transitions, is therefore required and could lead to significant improvements of the harvesting efficiency by constraining the system to a more efficient flapping.

\subsubsection{Effect of the mass ratio}
\begin{figure}
\begin{center}
\begin{tabular}{cc}
\includegraphics[width=.45\textwidth]{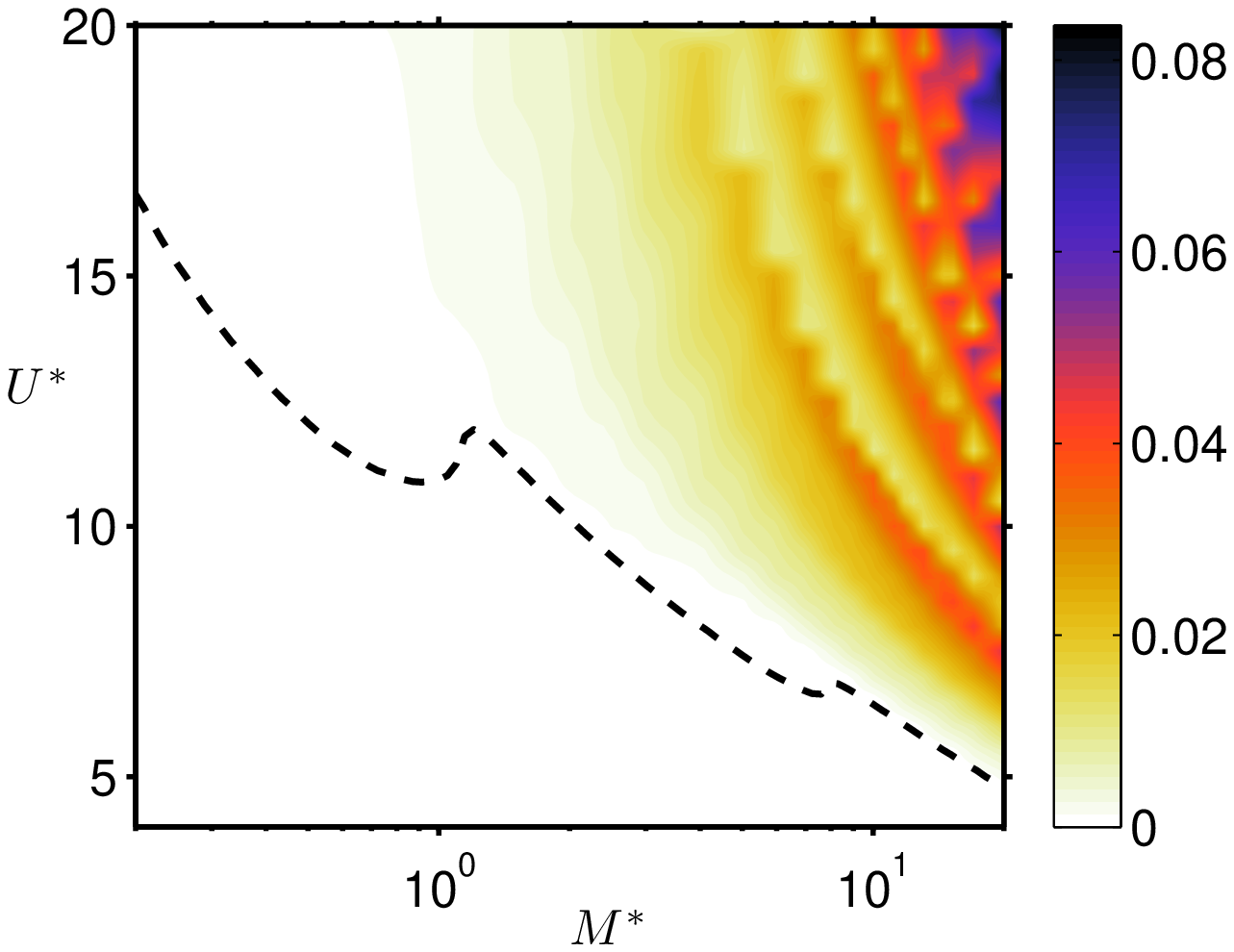} &
\includegraphics[width=.45\textwidth]{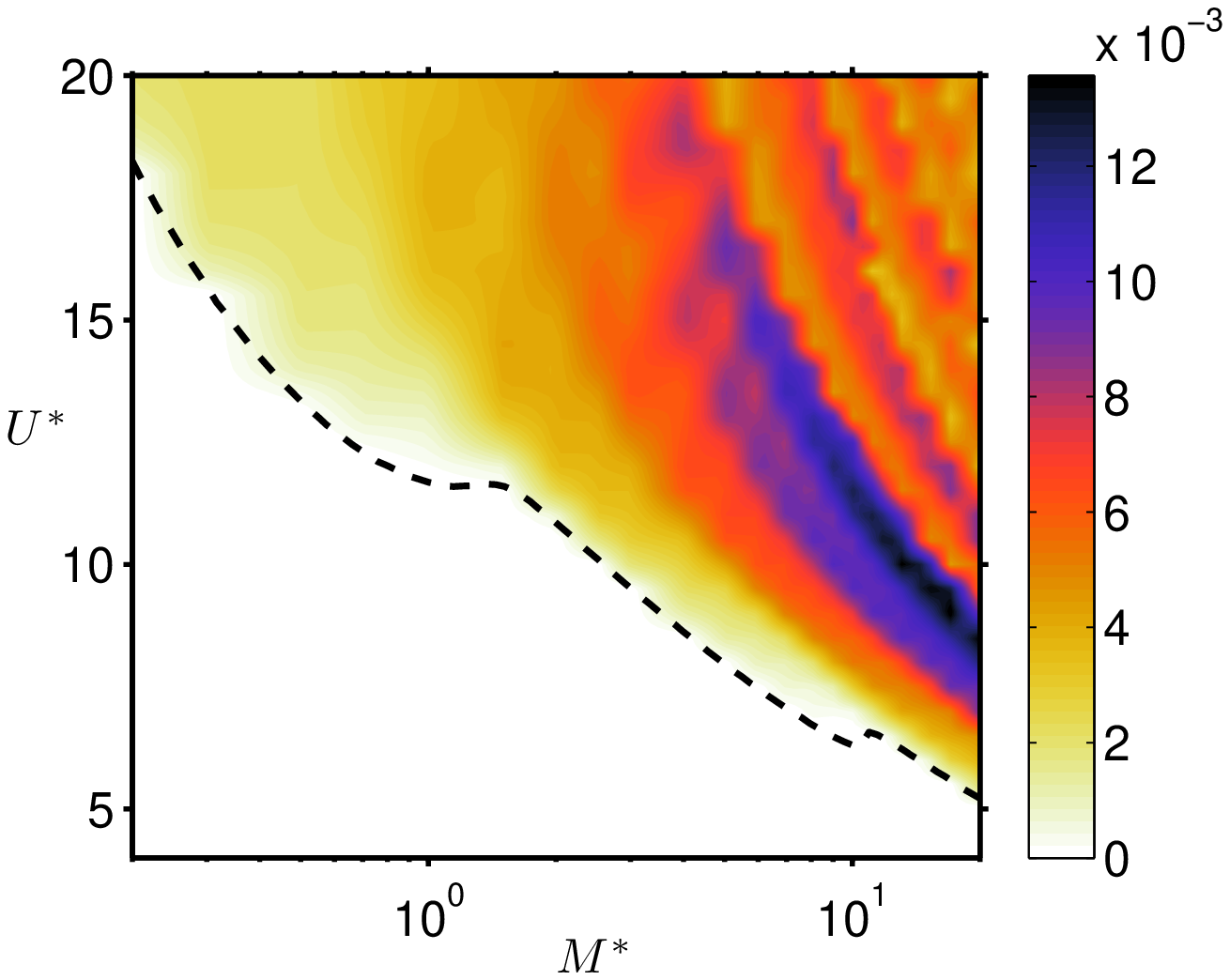}
\end{tabular}
\caption{Harvesting efficiency $\eta$ as a function of the mass ratio $M^*$ and normalized velocity $U^*$ for $\beta=0.1$ (left) and $\beta=1$ (right). In both cases, $\alpha=0.5$. The black dashed line corresponds to the instability threshold $U_c^*$ below which $\eta=0$.}\label{fig:eff_scanMU}
\end{center}
\end{figure}
The linear analysis of \citet{doare2011} identified significant differences in the performance of lighter (large $M^*$) or heavier flags (small $M^*$), as measured by the energy transfer from the structure to the output circuit. Higher performance at large $M^*$ was associated with the destabilization by damping of negative energy waves. 

\begin{figure}
\begin{center}
\includegraphics[width=.55\textwidth]{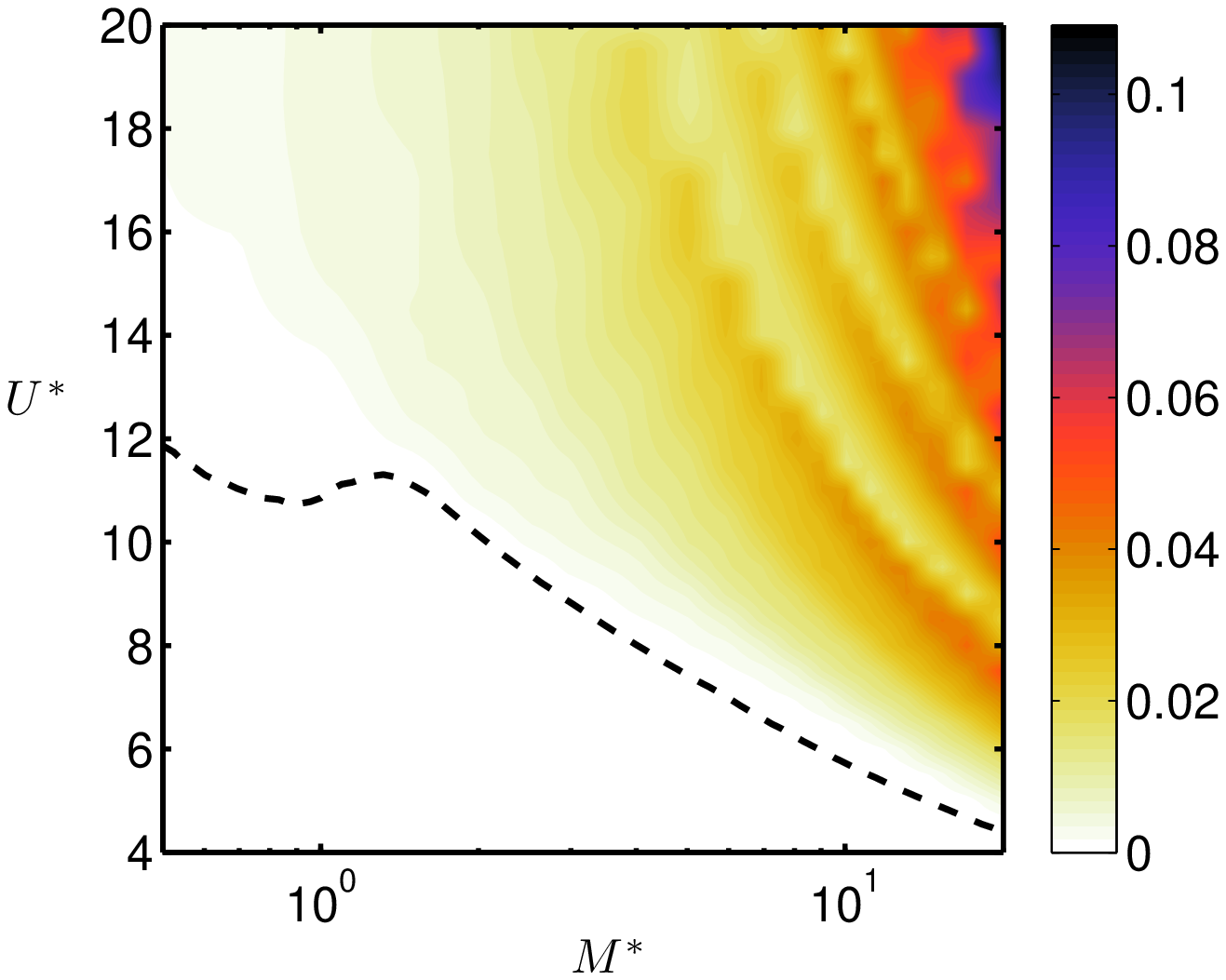}
\caption{Harvesting efficiency obtained for the optimal tuning $\beta$ as a function of the mass ratio $M^*$ and non-dimensional velocity $U^*$ for $\alpha=0.5$. For each value of $M^*$ and $U^*$, $\beta$ is chosen so as to maximize the harvesting efficiency. The black dashed line shows the minimum of the instability threshold over all possible values of $\beta$, below which $\eta=0$ for all $\beta$.}\label{fig:optim_MU}
\end{center}
\end{figure}

A similar result is observed here in nonlinear simulations for the harvesting efficiency $\eta$ (Figures \ref{fig:eff_scanMU} and \ref{fig:optim_MU}): harvesting efficiencies up to $10$--$12\%$ can be achieved for $M^*=20$ and $U^*\leq 20$, while the optimal value of $\eta$ is less than $1\%$ below $M^*=1$.

Comparing Figures \ref{fig:eff_scanMU}(a) and (b), the optimal $M^*$ appears as closely related to the tuning parameter $\beta$, emphasizing again the importance of the synchronization of the fluid-solid and electric systems: for small $\beta$, regions of greater $M^*$ will be optimal as they correspond to larger flapping frequencies while heavier flags (smaller $M^*$) will be optimal for larger values of $\beta$.

Finally, Figure \ref{fig:optim_MU} shows the optimal-tuning efficiency as a function of ($M^*$,$U^*$). Up to $12\%$ of the kinetic energy flux can be harvested for the largest value of $M^*$ and $U^*$ considered. However, it is also important to emphasize that this parameter region corresponds to closely-spaced mode switching events, making the efficiency of the system quite sensitive to fluctuations in the flow velocity.

\subsubsection{Effect of the piezoelectric coupling}
\begin{figure}
\begin{center}
\includegraphics[width=.55\textwidth]{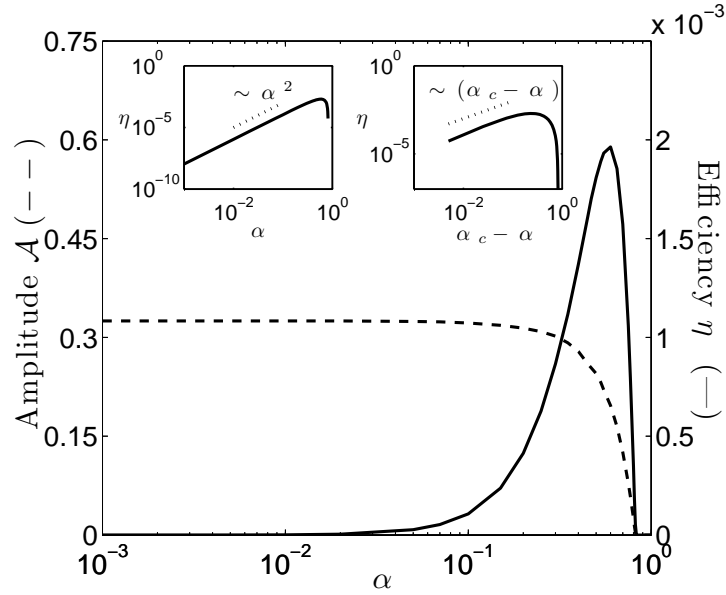}
\caption{Evolution of the flapping amplitude $\mathcal{A}$ (dashed) and harvesting efficiency $\eta$ (solid) with the coupling coefficient $\alpha$ for $\beta=1$, $M^*=0.5$ and $U^*=15$. The inset show the efficiency's scaling with $\alpha$ in the limit of $\alpha\ll 1$ and $\alpha_c-\alpha\ll 1$ with $\alpha_c\approx 0.82$ the critical value of $\alpha$ leading to restabilization of the piezoelectric flag for those parameter values. }\label{fig:alpha}
\end{center}
\end{figure}

The coupling coefficient $\alpha$ is a measure of the intensity of the fluid-solid and electric systems' forcing on each other, and as such is clearly expected to impact the amount of energy transferred to the output load. Figure \ref{fig:alpha} shows the evolution of $\mathcal{A}$ and $\eta$ when $\alpha$ is increased. For small coupling $\alpha\ll 1$, the flapping dynamics is only marginally modified and the amplitude of the charge transfer $q$ and electric potential $v$ increase linearly with $\alpha$ as seen in Eq.~\eqref{eq:piezond}. As a result, $\mathcal{Q}$ and $\eta$ initially increase quadratically with $\alpha$ (see inset on Figure \ref{fig:alpha}). However, when $\alpha$ is increased further the feedback piezoelectric coupling modifies the flapping dynamics resulting in a linear decrease of the flapping amplitude and harvesting efficiency and, eventually, the restabilization of the system. One can therefore identify an optimal value of the coupling coefficient, in the same way that an optimal damping was determined for maximum energy dissipation in \citet{singh2012b,singh2012}. The value of the optimal coefficient $\alpha_c$ clearly depends on the other system parameters, and will be greater when the flag is far from its stability threshold before piezoelectric coupling is introduced, or when destabilization by damping occurs as for larger $M^*$.

Achieving the optimal $\alpha$ is however not necessarily possible practically: $\alpha$ is a characteristic of the material's electric and mechanical properties and is of the order $\alpha\approx 0.3$ for typical piezoelectric materials such as PZT and even lower for PVDF \citep{doare2011}. Except in the vicinity of the instability threshold, the optimal $\alpha$ leading to maximum energy efficiency is however expected to be greater than this value, suggesting that an optimization of the piezoelectric flag design or future technical improvements in the properties of available piezoelectric materials can potentially increase the achievable values of $\alpha$ and lead to significant efficiency gains.

\section{Discussion and Perspectives}
\label{sec:conclusions}
The present study focused on the fully-coupled dynamics of a classical fluid-solid system, a flexible plate in axial flow, and a simple resistive circuit coupled through piezoelectric patches attached to the surface of the plate and converting the plate's bending deformation into an electric current. In the limit of continuous coverage by infinitesimal patches, the energy harvesting efficiency was determined as a function of the different system parameters, namely the inertia ratio, the non-dimensional flow velocity, the coupling coefficient and the tuning ratio. For realistic coupling coefficients, as much as $10\%$ of the kinetic energy flux can be transmitted to the output circuit, but this efficiency was found to be highly sensitive to several important parameters, in particular the coupling coefficient and the flow velocity. 

This study confirms the results by \citet{doare2011} on the impact of destabilization by damping and on the solid-electric energy transfers: in nonlinear saturated regimes, those parameter regions correspond indeed to maximum energy harvesting efficiency. The critical role played by the tuning ratio is also confirmed: maximum energy transfers are obtained when the output circuit characteristic timescale is tuned to the flapping frequency. This frequency is itself determined through the nonlinear coupling of the fluid, solid and electric systems, and modifications in the flapping frequency associated with a switch in the flapping mode directly impact the efficiency of the system and its robustness to fluctuations in flow velocity for example. Controlling the flapping mode selection is therefore an important challenge for the improvement of the efficiency of this model energy harvester, and should be considered in future work.

By coupling the fluid, solid and electric systems in a nonlinear model and by using an explicit description of the coupling mechanism and output circuit, the present approach provides some important insight on the nature and importance of the feedback of energy harvesting on the solid dynamics, as illustrated by the modification of the flapping amplitude and frequency, for example.  Even though the simplest possible circuit (a purely resistive element) was used here, the impact of the tuning ratio $\beta$ on the efficiency suggests that significant efficiency gains should be expected through a careful design of the output circuit, using more complex and possibly active circuits, as well as state-of-the-art power electronics techniques such as synchronized switching techniques \citep{lefeuvre2006}.

\begin{acknowledgements}
S. M. acknowledges the support of a Marie Curie International Reintegration Grant within the 7th European Community Framework Program (PIRG08-GA-2010-276762).
\end{acknowledgements}

\end{document}